\documentclass[12pt]{article} 

\usepackage{amssymb}
\usepackage{wasysym}
\newtheorem{lemma}{Lemma}
\newtheorem{theorem}{Theorem}

\newcommand{\ers}{\textsf{\textsc{EvenRankSum}}}
\newcommand{\mingap}{\textsf{\textsc{MinGap}}}

\newcommand{\qed}{\rule{0.5em}{1.5ex}}
\newcommand{\fqed}{{\hfill~\qed}}
\newenvironment{proof}{{\noindent \textbf{Proof.}}}
                      {{\hfill \fqed} \vspace{1em}}

\title{An $\Omega( n \log n)$ lower bound for computing the sum 
       of even-ranked elements}
\author{
Marc M\"orig\thanks{Faculty of Computer Science, 
                    University of Magdeburg, Magdeburg, Germany.}
\and
Dieter Rautenbach\thanks{Faculty of Mathematics and Natural Sciences,
                         Ilmenau University of Technology, Ilmenau, Germany.}
\and
Michiel Smid\thanks{School of Computer Science, Carleton University, 
                    Ottawa, Ontario, Canada. Research supported 
                    by NSERC.}
\and
Jan Tusch\footnotemark[1] 
}
\date{\today}

\begin{document}

\maketitle

\begin{abstract}
Given a sequence $A$ of $2n$ real numbers, the \ers\ problem asks for 
the sum of the $n$ values that are at the even positions in the sorted 
order of the elements in $A$. We prove that, in the algebraic 
computation-tree model, this problem has time complexity 
$\Theta(n \log n)$. This solves an open problem posed by 
Michael Shamos at the Canadian Conference on Computational Geometry 
in 2008. 
\end{abstract}

\section{Introduction} 
Let $A=(a_1,a_2,\ldots,a_{2n})$ be a sequence of $2n$ real numbers. 
We define the \emph{even-rank-sum} of $A$ to be the sum of the $n$ 
values that are at the even positions in the sorted order of the 
elements in $A$. Formally, let $\pi$ be a permutation of 
$\{1,2,\ldots,2n\}$ that sorts the sequence $A$ in non-decreasing 
order; thus, $a_{\pi(1)} \leq a_{\pi(2)} \leq \ldots \leq a_{\pi(2n)}$. 
Then the even-rank-sum of the sequence $A$ is the real number 
\[ a_{\pi(2)} + a_{\pi(4)} + a_{\pi(6)} + \ldots +  a_{\pi(2n)} . 
\] 
Observe that any permutation $\pi$ that sorts the sequence $A$ in 
non-decreasing order gives rise to the same even-rank-sum. 
We consider the following problem: 

\vspace{0.5em} 
 
\noindent 
\ers: Given a sequence $A$ of $2n$ real numbers, compute the 
even-rank-sum of $A$. 

\vspace{0.5em} 

By using an $O(n \log n)$--time sorting algorithm, this problem can 
be solved in $O(n \log n)$ time. In the Open Problem Session at the 
Canadian Conference on Computational Geometry in 2008, 
Michael Shamos posed the problem of proving an $\Omega(n \log n)$ 
lower bound on the time complexity of \ers~ in the algebraic 
computation-tree model. (See \cite{b-lbact-83,ps-cgi-88} for a 
description of this model.) In this paper, we present such a proof: 

\begin{theorem}  \label{thm1}  
       In the algebraic computation-tree model, the time complexity 
       of \ers\ is $\Theta(n \log n)$. 
\end{theorem} 

We prove Theorem~\ref{thm1} by presenting an $O(n)$--time reduction of 
\mingap\ to \ers. The former problem is defined as follows.   
Let $X=(x_1,x_2,\ldots,x_n)$ be a sequence of $n$ real numbers, and 
let $\pi$ be a permutation of $\{1,2,\ldots,n\}$ such that 
$x_{\pi(1)} \leq x_{\pi(2)} \leq \ldots \leq x_{\pi(n)}$. 
For each $1 \leq i < n$, we define the difference
$x_{\pi(i+1)} - x_{\pi(i)}$ to be a \emph{gap} in the sequence $X$.

\vspace{0.5em} 

\noindent 
\mingap: Given a sequence $X=(x_1,x_2,\ldots,x_n)$ of $n$ real numbers 
and a real number $g>0$, decide if each of the $n-1$ gaps in $X$ is 
at least $g$. 

\vspace{0.5em} 

Since in the algebraic computation-tree model, \mingap\ has an 
$\Omega(n \log n)$ lower bound (see~\cite[Section~8.4]{ps-cgi-88}), 
our reduction will prove Theorem~\ref{thm1}.  

\section{The proof of Theorem~\ref{thm1}} 
We now show how to reduce, in $O(n)$ time, \mingap\ to \ers.

Let $\mathcal{A}$ be an arbitrary algorithm that solves \ers. 
We show how to use algorithm $\mathcal{A}$ to solve \mingap.  
Let $n \geq 2$ be an integer and consider a sequence 
$X=(x_1,x_2,\ldots,x_n)$ of $n$ real numbers and a real number $g>0$. 
The algorithm for solving \mingap\ makes the following three steps:

\vspace{0.5em} 

\noindent 
\textbf{Step 1:} Compute $S = \sum_{i=1}^n x_i$ and, for 
$i=1,2,\ldots,n$, compute $a_{2i-1} = x_i$ and $a_{2i} = x_i + g$.

\vspace{0.5em} 

\noindent 
\textbf{Step 2:} Run algorithm $\mathcal{A}$ on the sequence 
%% A=$(a_1,a_2,\ldots,a_{2n})$,  
%% MS weil A nicht benutzt wird, habe ich es weggelassen  
$(a_1,a_2,\ldots,a_{2n})$, 
and let $R$ be the output, i.e., $R$ is the even-rank-sum of this 
sequence.

\vspace{0.5em} 

\noindent 
\textbf{Step 3:} If $R = S + ng$, then return YES. Otherwise, return NO. 

\vspace{0.5em} 

It is clear that the running time of this algorithm is $O(n)$ plus the 
running time of $\mathcal{A}$. Thus, it remains to show that the 
algorithm correctly solves \mingap. 
%% That is, to show 
%% MS 
That is, we have to show 
that the minimum gap $G$ of $X$ is at least $g$ if and only if 
$R = S + ng$. 
%% which is an immediate consequence of the following lemma:
%% MS 
This is an immediate consequence of the following lemma:
\begin{lemma}
\label{lem1}
Let $x_1,x_2,\ldots,x_n$ and $g$ be real numbers such that $x_1\leq x_2\leq \ldots \leq x_n$ and $g>0$.
Let $(a_1,a_2,\ldots,a_{2n})=(x_1,x_1+g,x_2,x_2+g,\ldots,x_n,x_n+g)$
and let $\pi$ be a permutation of $\{1,\ldots,2n\}$ such that
$b_1\leq b_2\leq \ldots \leq b_{2n}$
with $b_i=a_{\pi(i)}$ for $1\leq i\leq 2n$.

If $R=\sum\limits_{i=1}^n b_{2i}$,
$U=\sum\limits_{i=1}^n b_{2i-1}$, and
$G=\min\{ x_{i+1}-x_i\mid 1\leq i\leq n-1\}$,
then
$R-U\leq ng$ with equality if and only if $G\geq g$.
\end{lemma}
\begin{proof}
Since $x_1,x_1+g,x_2,x_2+g,\ldots,x_i,x_i+g\leq x_i+g$,
we have $x_i+g\geq b_{2i}$ for $1\leq i\leq n$.
Since $x_i,x_i+g,x_{i+1},x_{i+1}+g,\ldots,x_n,x_n+g\geq x_i$,
we have $x_i\leq b_{2i-1}$ for $1\leq i\leq n$.
Hence $b_{2i}-b_{2i-1}\leq (x_i+g)-x_i=g$ for $1\leq i\leq n$
which implies $R-U\leq ng$.

If $G\geq g$, then clearly $R-U=ng$.
Conversely, if $R-U=ng$, then $b_{2i}-b_{2i-1}=g$ for $1\leq i\leq n$.
In view of the above, this implies that
$x_i+g=b_{2i}$ and $x_i=b_{2i-1}$ for $1\leq i\leq n$.
Since $x_{i+1}=b_{2i+1}\geq b_{2i}=x_i+g$ for $1\leq i\leq n-1$,
we obtain $G\geq g$.
\end{proof}

We complete the proof of Theorem~\ref{thm1} by observing that $R+U=2S+ng$ and by Lemma~\ref{lem1} we have $G \ge g$ if and only if $R=U+ng=S+ng$.

\bibliographystyle{plain}
\bibliography{evenranksum}

\end{document}